\documentclass[journal]{IEEEtran}
\ifCLASSINFOpdf
\else
\fi

\usepackage{stfloats}
\usepackage[fleqn]{amsmath}

\usepackage{graphicx,epsfig,color}

\usepackage{enumerate}
\usepackage[noend]{algpseudocode}
\usepackage{algorithmicx,algorithm} 

\newcounter{savenumi}

\newtheorem{theoremfoo}{Theorem}

\newtheorem{propositionfoo}[theoremfoo]{Proposition}

\newtheorem{lemmafoo}[theoremfoo]{Lemma}

\newtheorem{conjecturefoo}[theoremfoo]{Conjecture}

\newtheorem{corollaryfoo}[theoremfoo]{Corollary}

\newtheorem{exercisefoo}{Exercise}

\newtheorem{openfoo}[theoremfoo]{Question}

\newtheorem{nttn}[theoremfoo]{Notation}

\newtheorem{dfntn}[theoremfoo]{Definition}

\renewcommand{\theenumi}{\roman{enumi}}

\renewcommand{\tt}{{\rm tt}}    
\def\nre.{$n$\/-r.e.}

\newtheorem{factfoo}[theoremfoo]{Fact}

\newcommand{\squeeze}{
\textwidth 6in
\textheight 8.8in
\oddsidemargin 0.2in
\topmargin -0.4in
}

\newtheorem{propertyfoo}[theoremfoo]{Property}

\makeatletter

\def\@makechapterhead#1{ \vspace*{50pt} { \parindent 0pt \raggedright 
 \ifnum \c@secnumdepth >\m@ne \huge\bf \@chapapp{} \thechapter. \par 
 \vskip 20pt \fi \Huge \bf #1\par 
 \nobreak \vskip 40pt } }

\def\@sect#1#2#3#4#5#6[#7]#8{\ifnum #2>\c@secnumdepth
     \def\@svsec{}\else 
     \refstepcounter{#1}\edef\@svsec{\csname the#1\endcsname.\hskip 1em }\fi
     \@tempskipa #5\relax
      \ifdim \@tempskipa>\z@ 
        \begingroup #6\relax
          \@hangfrom{\hskip #3\relax\@svsec}{\interlinepenalty \@M #8\par}
        \endgroup
       \csname #1mark\endcsname{#7}\addcontentsline
         {toc}{#1}{\ifnum #2>\c@secnumdepth \else
                      \protect\numberline{\csname the#1\endcsname}\fi
                    #7}\else
        \def\@svsechd{#6\hskip #3\@svsec #8\csname #1mark\endcsname
                      {#7}\addcontentsline
                           {toc}{#1}{\ifnum #2>\c@secnumdepth \else
                             \protect\numberline{\csname the#1\endcsname}\fi
                       #7}}\fi
     \@xsect{#5}}

\def\@begintheorem#1#2{\it \trivlist \item[\hskip \labelsep{\bf #1\ #2.}]}

\def\@opargbegintheorem#1#2#3{\it \trivlist
      \item[\hskip \labelsep{\bf #1\ #2\ (#3).}]}

\makeatother

\newif\ifshortconferences
\shortconferencesfalse
\newif\ifmediumconferences
\mediumconferencesfalse

\def\ending#1{{\count1=#1\relax
\count2=\count1
\divide\count2 by 100
\multiply\count2 by 100
\advance\count1 by -\count2
\ifnum\count1=11
th%
\else \ifnum\count1=12
th%
\else \ifnum\count1=13
th%
\else 
\count2=\count1
\divide\count1 by 10
\multiply\count1 by 10
\advance\count2 by -\count1
\ifnum\count2=1
st%
\else \ifnum\count2=2
nd%
\else \ifnum\count2=3
rd%
\else th%
\fi\fi\fi\fi\fi\fi
}}

\def\Proceedingsofthe{\ifshortconferences Proc.\else\ifmediumconferences Proc.\else Proceedings of the\fi\fi}

\newcounter{confnum}

\def\conf#1#2{%
\setcounter{confnum}{#2}%
\addtocounter{confnum}{-\csname #1zero\endcsname}%
\ifnum\value{confnum}=1%
\expandafter\ifx\csname #1One\endcsname\relax%
\Proceedingsofthe\ \arabic{confnum}\ending{\value{confnum}}\ \csname #1name\endcsname%
\else \csname #1One\endcsname\fi%
\else%
\Proceedingsofthe\
\arabic{confnum}\ending{\value{confnum}}\ \csname #1name\endcsname\fi}

\def\qsym{\vrule width0.7ex height0.9em depth0ex}

\newif\ifqed\qedtrue

\def\noqed{\global\qedfalse}

\def\qed{\ifqed{\penalty1000\unskip\nobreak\hfil\penalty50
\hskip2em\hbox{}\nobreak\hfil\qsym
\parfillskip=0pt \finalhyphendemerits=0\par\medskip}\fi\global\qedtrue}

\makeatletter
\def\eqnqed{\noqed
	\def\@tempa{equation}
	\ifx\@tempa\@currenvir\def\@eqnnum{\qsym}%
	\addtocounter{equation}{-1}\else%
    \def\@@eqncr{\let\@tempa\relax
    \ifcase\@eqcnt \def\@tempa{& & &}\or \def\@tempa{& &}%
      \else \def\@tempa{&}\fi
     \@tempa {\def\@eqnnum{{\qsym}}\@eqnnum}
     \global\@eqnswtrue\global\@eqcnt\z@\cr}\fi}

\def\eqnlabel#1#2{\if@filesw {\let\thepage\relax%
   \def\protect{\noexpand\noexpand\noexpand}%
   \edef\@tempa{\write\@auxout{\string
      \newlabel{#2}{{{#1}}{\thepage}}}}%
   \expandafter}\@tempa%
   \if@nobreak \ifvmode\nobreak\fi\fi\fi%
	\def\@tempa{equation}
	\ifx\@tempa\@currenvir\def\theequation{{#1}}%
	\addtocounter{equation}{-1}\else%
    \def\@@eqncr{\let\@tempa\relax
    \ifcase\@eqcnt \def\@tempa{& & &}\or \def\@tempa{& &}%
      \else \def\@tempa{&}\fi
     \@tempa {\def\@eqnnum{{#1}}\@eqnnum}
     \global\@eqnswtrue\global\@eqcnt\z@\cr}\fi}

\makeatother

\def\QED{\qed}

\makeatother

\squeeze

\newcommand{\format}{png}

\begin{document}
%
\title{Edge computing based incentivizing mechanism for mobile blockchain in IOT}
%
%
%
\author {Liya Xu, Mingzhu Ge$^*$, Weili Wu, ~\IEEEmembership{Member,~IEEE}\\

\thanks{Liya Xu, School of Information Science and Technology, Jiujiang University, jiujiang, 332005, China, e-mail: xuliya603@whu.edu.cn}
\thanks{Mingzhu Ge, Department of Information Technology Center, Jiujiang University, Jiujiang, 332005, China, e-mail: mingzhug1989@gmail.com}
\thanks{Weili Wu, Department of Computer Science, University of Texas at Dallas, Richardson, TX 75080, USA, e-mail: weiliwu@utdallas.edu}

}

\maketitle

\begin{abstract}
Mining in the blockchain requires high computing power to solve the hash puzzle for example proof-of-work puzzle. It takes high cost to achieve the calculation of this problem in devices of IOT, especially the mobile devices of IOT. It consequently restricts the application of blockchain in mobile environment. However, edge computing can be utilized to solve the problem for insufficient computing power of mobile devices in IOT. Edge servers can recruit many mobile devices to contribute computing power together to mining and share the reward of mining with these recruited mobile devices. In this paper, we propose an incentivizing mechanism based on edge computing for mobile blockchain. We design a two-stage Stackelberg Game to jointly optimize the reward of edge servers and recruited mobile devices. The edge server as the leader sets the expected fee for the recruited mobile devices in Stage I. The mobile device as a follower provides its computing power to mine according to the expected fee in Stage. It proves that this game can obtain a uniqueness Nash Equilibrium solution under the same or different expected fee. In the simulation experiment, we obtain a result curve of the profit for the edge server with the different ratio between the computing power from the edge server and mobile devices. In addition, the proposed scheme has been compared with the MDG scheme for the profit of the edge server. The experimental results show that the profit of the proposed scheme is more than that of the MDG scheme under the same total computing power.
\end{abstract}


%
\IEEEpeerreviewmaketitle

\section{Introduction}
%
%
%
%
\par Online payment has become an indispensable way of payment in people's life. In order to ensure the security of payment, online payment is achieved through an authoritative agent. However, the cost of establishing an authority is enormous, which increases the cost of payment. Blockchain technology can realize P2P payment\cite{aste2017blockchain}. Blockchain is a distributed ledger that is copied synchronously across multiple users, which differs from the traditional centralized ledger, and Bitcoin is used to solve the dual payment problem\cite{tschorsch2016bitcoin}. Blockchain has surpassed its original design and become the basic technology to realize decentralized control. Compared with centralized control, the blockchain system has the advantages of distributed high redundancy storage, time sequence data, tamper-resistant and forgery, decentralized credit, automatic execution of smart contracts, security and privacy protection.

\par The key process of blockchain is a calculation process to solve the hash puzzle, which requires high computing power. However, the mobile devices in IOT have insufficient computing power. Therefore, edge computing has been paid more and more attention of scholars\cite{satyanarayanan2017emergence}\cite{luo2020edge}. The target of edge computing is to enable a large number of devices to run applications on IOT, which enables these devices to provide their computing power for mining\cite{yu2017survey}. Therefore, it is a trend that edge computing is used to solve the hash puzzle in blockchain\cite{yeow2017decentralized}. By combining blockchain with edge computing, edge servers can organize a large number of devices to provide computing power, storage resources, etc. It can greatly improve the transmission efficiency of the system and ensure data integrity and computing effectiveness. With the participation of edge computing, the blockchain system obtains a large number of computing power on IOT, which reduces the burden of computing power limited devices for mining. It enables out of chain storage and out of chain computing of edge computing to be realized, and satisfies scalable storage and computing on the blockchain\cite{liu2017blockchain}.

\par However, blockchain is not widely used in mobile environment. Due to the limited computing power of mobile devices, it costs too much to solve the POW puzzle. Therefore, we need to further research the mining strategy in the mobile environment to promote the application of blockchain in IOT. The incentive mechanism based on edge computing is considered to overcome the lack of computing power in mobile devices and satisfies the requirement of mining. Edge servers act as miner, but the number is small, while the number of mobile devices is large. Therefore, the system encourages the edge server to actively recruit mobile devices to provide computing power to complete the mining task. When the mining is successful, the edge server and the mobile device share the rewards. The contributions of this work are mainly shown as follows.
\begin{enumerate}
	\item We discuss the challenges of mobile blockchain in mining. In order to address the challenge for high cost of mining by mobile devices alone, the edge server recruit mobile devices to provide computing power to mine and share the profit with them. In order to make them obtain the maximum profit, this paper formulates a two-stage Stackelberg game model to set the profit distribution of mining.
	\item We adopt backward induction to prove that the uniqueness Nash equilibrium solution exists in this game under the same expected fee and different expected fee.
	\item We show the result curve of the profit for the edge server with the different ratio between the computing power from the edge server and mobile devices. The results show that the contribution of the computing power from the edge server to its profit is more than that of it recruited mobile devices under the same condition.
\end{enumerate}

\par The rest of this paper is organized as follows. Related work is briefly reviewed in Section~\ref{related-work-section}. In Section ~\ref{model-section}, we give out the detail of system model and problem formulation. In Section ~\ref{Game-Equilibrium-Analysis}, Game Equilibrium is analyzed. Performance evaluation and analysis are given in Section ~\ref{simlation-section}. The last Section is conclusion.

\section{Related work}\label{related-work-section}

\par  In recent years, the technology of integrating blockchain into edge computing has gradually attracted the attention of scholars. In \cite{khan2019blockchain}, the authors designed an architecture to combine blockchain technology with edge computing. It achieves the fine management of different department data. They implement the synchronization of data storage and processing by deploying regional blockchain, and reach an agreement through sharing. In \cite{ damianou2019architecture}, the author overcomes the constraints of the Internet of things by combining edge computing and blockchain. It can ensure the security and privacy of Internet of things devices when it translates sensitive personal data. Therefore, it can make IOT devices get rid of the dependence on storage devices and improve the overall performance. In \cite{zhang2019edge}, the author designs an architecture combining edge intelligence with blockchain authorization to achieve efficient edge service management. For edge resource sharing, they designed a cross domain scheduling mechanism, and used the credit approval mechanism to ensure security, which can reduce service costs and improve service capabilities.

\par Many researches have been carried out on the mining scheme of blockchain by formulating game model. Houy \cite{houy2014bitcoin}proposed a noncooperative game model in the mining process. In this model, the events of POW puzzle follow Poisson distribution. When the equilibrium point of Nash is reached, the optimal solution of both sides is obtained, but only the equilibrium solution between two miners is obtained. Kiayias et al \cite{kiayias2016blockchain}considered the mining of miners as a competitive game process. Assuming that the miners are rational, they compete for mining and choose whether to broadcast their solutions based on their profit. It is found that under some conditions, the optimal solution is not to propagate. In addition, the authors proved that there is multiplicity of Nash equilibrium in this model. Similar to \cite{kiayias2016blockchain}, L.Wang and Y. Liu \cite{wang2015exploring} considered the mining of miners as a random game model. The miners decide whether to propagate the block according to the estimation of reward.

\par  The traditional mining of miners in the blockchain is carried out individually. The advantage of mining by miners alone is that when miners successfully calculate the hash value that meets the conditions, they can obtain all rewards. However, this mining scheme is inefficient and has unstable earnings. In order to obtain stable rewards, researchers have introduced mine pools, which is another way to concentrate resources to mining blocks\cite{ beccuti2017bitcoin}. Lewenberg et al \cite{lewenberg2015bitcoin} proposed the mining scheme of blockchain based on cooperative game. They use the theoretical tool of cooperative game to research the mining pool that miners would like to join and reasonably shared their rewards. The interaction between the miner and the mining pool is simulated as a joint game. In order to the data processing in the Internet of things authorized by the blockchain, \cite{chen2019cooperative} proposed a multi-hop cooperative distributed algorithm to achieve the mining tasks. They set the competition among devices as a game problem to reduce the computing cost of Internet of things devices. Each device can decide its own computing power to obtain the maximum benefit. The researchers found that when the scheme is applied to bitcoin network, some miners always turn to other mines for higher expected rewards under any incentive allocation scheme

\par However, the above research works are to use dedicated nodes to mine, without considering the application of mobile environment. In the mobile Internet environment, the computing power of a single mobile smart device is often limited, but the number of these devices is large. Therefore, it is a new opportunity for blockchain application that edge servers organize a large number of mobile devices, and make use of their idle computing power to contribute computing power to mining.

\section{System Model and Problem Formulation}\label{model-section}

\subsection{System Model}

\par The process of consensus management contains these steps as following: Block broadcast, Mining, Progagation, 
Verification, Confirm and Add new block. The data block that contains transaction records is periodically broadcasted
 in blockchain network. These transaction records are the issued transactions. The miners will solve a crypto-puzzle by 
 theirs computing power according to given parameters of the blockchain \cite{li2017securing}. If the miner successfully solves the given puzzle, it propagates the result and block to other miners in the blockchain system to verify. The fastest miner who solves the puzzle and passes the verification will obtain the reward R of blockchain system. To finish the computation and propagation as soon as possible, the edge servers recruit the mobile devices to compute beside theirselves. The verification tasks are completely handed over to the mobile devices. Each mobile device is assigned the computation and verification task according to its computing power by edge servers and then get reward from the edge server that recruited it. Each edge server should share the given reward from the blockchain to mobile devices based on their contributions. In addition, the calculation and verification of blocks will generate communication overhead.

\subsection{Problem Formulation}

\par Let $\mathnormal E=\left\{E_1, \cdots, E_x\right\}$ represents a group of edge servers. Each edge server recruits mobile devices(denoted by $\mathcal M=\left\{M_1, \cdots, M_y\right\}$) to provide computation and verification services. The miners compete with each other to mine. The fastest miner who solves the puzzle and passes the verification will obtain the reward of blockchain system. The reward consists of fixed reward $R$ and variable transaction reward $TR$. For miner $E_i$, the probability of success in mining is determined by its computing power, which is represented by $\Pr=\alpha e^{-\gamma zT}$ \cite{houy2014bitcoin}, where $\alpha$ is factor about the computing power. 
\begin{eqnarray}
\alpha_i(x_i,\mathcal M_{-1})=\frac{x_i}{\sum_{j \in \mathcal M}x_j},\alpha_i>0
\end{eqnarray}
and $\sum_{j \in \mathcal M}\alpha_j=1$. The process of solving puzzle follows Poisson distribution, and $\gamma $ is Poisson parameter\cite{xiong2017edge}. $z>0$ represents a delay factor in transmission, and $T$ represents the number of transactions in one block. 

The profit of the edge server $E_i$ contain the three parts as follow: 1) The fixed reward from mining; 2)the cost for mobile device charging; 3)electricity and other costs. The utility function of edge servers is formulated as follow: 

\begin{eqnarray}
U_e=(R+TR) e^{-\gamma zT}-\sum_{i=1}^n f_i-\Phi.\label{utility of edge}
\end{eqnarray}

The profit of mobile devices contain the three parts as follow: 1)The expected fee provided by edge server; 2)mined block verification; 3)electricity and other costs. The profit function of mobile devices is formulated as follow:

\begin{eqnarray}
U_m=f_i\frac{x_i}{\sum_{j \in \mathcal M}x_j}e^{-\gamma zT_m}-\varphi x_i.\label{utility of miner i}
\end{eqnarray}

$f_i$ is the reward from edge server depend on its computing power.$\varphi$ is the resource about computing power, memory and so on.

The process of mining block between the edge server and mobile devices is designed as a two-stage Stackelberg game. The edge server acts as leader. It sets the expected fee for mobile devices in Stage \uppercase\expandafter{\romannumeral1}. The mobile devices act as follower. They provide their computing power for the leader in Stage \uppercase\expandafter{\romannumeral2}. Obviously, for followers, if the expected fee is less than their own consumption, they will not accept the recruitment of leaders. The minimum consumption of a mobile miner is defined as $\omega_{min}$. Therefore, the expected fee given by leader to each mobile miner must be greater than the consumption value of follower. Then the objective functions of leaders and followers can be expressed as follows:

\begin{eqnarray}
Leader:max F(x_i) \notag\\
1 \ge \alpha_i >0 \quad\quad\notag\\
Follower: max f(\omega)\notag\\
\omega > \omega_{min}\quad\quad 
\end{eqnarray} 


\section{Game Equilibrium Analysis}\label{Game-Equilibrium-Analysis}

In our model, we consider two situations for mobile devices as follow: the same computer power for each miner and the different computer power for each miner. The traditional backward induction\cite{xiong2017edge} will be adoped to analyze this game.

\subsection{The same computer power for each miner} \label{The same computer power for each miner}
 \emph{1) Stage \uppercase\expandafter{\romannumeral2}: Miner's Game}: we first analyze the simple situation that is the same computing power for each miner. These miners can be considered as a entirety, and their total computer power is $Y$. The computing power from the edge server is $X$. Then, the miners compete with the edge server to maximize their profit by providing their computer power, which forms the Edge Miner Game(EMG) $E_m=\{E,\mathcal M,U_m\}$. $E$ is the edge server, $\mathcal M$ is the group of mobile miners recruited by the edge server, $U_m$ represents the utility function of mobile miners. Therefore, the profit function of mobile miners, in Stage \uppercase\expandafter{\romannumeral2}, is represented as:

\begin{eqnarray}
U_m=P\frac{Y}{X+Y}e^{-\gamma zT_m}-\varphi_1 Y.\label{utility of total miners}
\end{eqnarray}
where $P$ is expected fee of miners from edge server.

{\bf Theorem 1.} There is a Nash equilibrium in $E_m=\{E,\mathcal M,U_m\}$ 
\par $proof.$ The computer power space of each miner is a continuous and non-empty value range. $U_m$ is obviously continuous with respect to $Y$. Let $\alpha=\frac{Y}{X+Y}$. Then, we prove the concavity of $U_m$. $U_m$ will be concave if it satisfy the two condition: the first derivative of $U_m$ with respect to $Y$ is positive and the second derivative of $U_m$ with respect to $Y$ is negative. The process is as following.

\begin{eqnarray}
\frac{\partial U_m}{\partial Y}=P \frac{\partial\alpha}{\partial Y}-\varphi_1 \label{the first order U_m}
\end{eqnarray}
and 
\begin{eqnarray}
\frac{\partial^2 U_m}{\partial Y^2}=P \frac{\partial^2\alpha}{\partial Y^2}
\end{eqnarray}
where
\begin{eqnarray}
\frac{\partial \alpha}{\partial Y}= \frac{X}{(X+Y)^2}>0
\end{eqnarray}
and
\begin{eqnarray}
\frac{\partial^2 \alpha}{\partial Y^2}= -\frac{2X}{(X+Y)^3}<0
\end{eqnarray}

Therefore, $U_m$ satisfies the concave condition with respect to $Y$. According to \cite{han2012game}, the Nash equilibrium is in the EMG $E_m$. 

Let  (\ref{the first order U_m})=0, we can get the best function of miners as following,
\begin{eqnarray}
\frac{\partial U_m}{\partial Y}=Pe^{-\gamma zT_m} \frac{\partial\alpha}{\partial Y}-\varphi_1 =0 \label{the first derivative equal 0 in I of miners}
\end{eqnarray}

\begin{eqnarray}
Y^*=F(x)=\sqrt{\frac{Pe^{-\gamma zT_m}X}{\varphi_1}}-X \label{unique 1 Nash}  
\end{eqnarray}

{\bf Theorem 2.}There is a unique Nash equilibrium in EMG $E_m$ if the following condition is satisfied
\begin{eqnarray}
X<\mathop{min}\limits_{P>0}\{\int \frac{Pe^{-\gamma zT_m}}{4\varphi_1} d_X, \frac{Pe^{-\gamma zT_m}}{\varphi_1}\} \label{condition 1}
\end{eqnarray}

\par Let $Y^*$ represent the Nash equilibrium of the EMG. If $Y^*=F(Y)$ is the standard function, then the EMG $E_m$ exists the unique Nash equilibrium\cite{han2012game}.

{\bf Definition 1.} A funciton $F(x)$ is a standard function if the following three conditions are satisfied\cite{han2012game}:
\par (1)$F(x)$ is positive.
\par (2)$F(x)$ is monotonous. It is formulated as: if $x \le x'$, then $F(x)\le F(x')$;
\par (3)$F(x)$ is scalable. It is formulated as: for all $\lambda>1$, then $\lambda F(x)>F(\lambda x)$;

Firstly, for $F(x)>0$, from (\ref{unique 1 Nash}), we have

\begin{eqnarray}
If X<\frac{Pe^{-\gamma zT_m}}{\varphi_1}, then \sqrt{\frac{Pe^{-\gamma zT_m}X}{\varphi_1}}-X>0   
\end{eqnarray}

Under the condition in (\ref{condition 1}), we can obtain
\begin{eqnarray}
\sqrt{\frac{Pe^{-\gamma zT_m}X}{\varphi_1}}-X>0
\end{eqnarray}

\begin{figure*}[b]
	\hrulefill 
	\begin{align*}
	F(x')-F(x) &=\sqrt{\frac{Pe^{-\gamma zT_m}X'}{\varphi_1}}-X'-(\sqrt{\frac{Pe^{-\gamma zT_m}X}{\varphi_1}}-X) \\
	&=\sqrt{\frac{Pe^{-\gamma zT_m}}{\varphi_1}}(\sqrt{X'}-\sqrt{X})-(X'-X) \\
	&= (\sqrt{\frac{Pe^{-\gamma zT_m}}{\varphi_1}}-\sqrt{X'}-\sqrt{X})(\sqrt{X'}-\sqrt{X})
	\tag{13}
	\end{align*}
	\begin{align*}
	\lambda F(x)-F(\lambda X) &=\lambda\sqrt{\frac{Pe^{-\gamma zT_m}X}{\varphi_1}}-\lambda X-(\sqrt{\frac{Pe^{-\gamma zT_m}\lambda X}{\varphi_1}}-\lambda X) \\
	&=(\lambda -\sqrt{\lambda})\sqrt{\frac{Pe^{-\gamma zT_m}X}{\varphi_1}}
	\tag{14}\label{for condition 3 of definition 1}
	\end{align*}
\end{figure*}

Secondly, Let $X' \ge X$, we calculate the formula of $F(X')-F(X)$, which is shown in (13). Obviously, $\sqrt{X'}-\sqrt{X} \ge 0$. In addition, 
\begin{eqnarray}
\setcounter{equation}{15}
\sqrt{\frac{Pe^{-\gamma zT_m}}{\varphi_1}}-\sqrt{X'}-\sqrt{X} \ge \notag\\ \sqrt{\frac{Pe^{-\gamma zT_m}}{\varphi_1}}-2\sqrt{X'}
\end{eqnarray} 

Under the condition in (\ref{condition 1}), we can prove that 
\begin{eqnarray}
\sqrt{\frac{Pe^{-\gamma zT_m}}{\varphi_1}}-2\sqrt{X'} \ge 0
\end{eqnarray}
Thus, $F'(X)-F(x) \ge 0$.

Finally, for condition(3) of Definition 1. We prove the formula of $\lambda F(x)>F(\lambda x)$ with respect to $\lambda>1$. The process of calculation is shown in (\ref{for condition 3 of definition 1}). 

\par Therefore, the three conditions shown in Definition 1 have been proved for the response function in (\ref{unique 1 Nash}). The process of proof has done..

\par From above, we can obtain that the uniqueness Nash equilibrium for miners is true in EMG $E_m$ when $Y^*=\sqrt{\frac{Pe^{-\gamma zT_m}X}{\varphi_1}}-X$. And now, we need to analyze the optimal profit of the edge server in the first stage.

\emph{2) Stage \uppercase\expandafter{\romannumeral1}: Edge server's optimal profit}: According to the Nash equilibrium of the computer power provided by mobile miners in EMG $E_m$, the Edge Server is as leader that can optimize its computer power to achieve its optimal profit represented in (\ref{utility of edge}). For the edge server, the computer power $X$ belongs to itself. The revenue generated by this part of computing power $X$ belongs to the edge server. What we need to analyze is the additional benefits of the computing power from the edge server recruiting mobile miners. The total computer power provided by miners is $Y$. Therefore, the additional benefits for the edge server represented as:

\begin{eqnarray}
\Delta U_e=(R+TR)\frac{Y}{X+Y}e^{-\gamma zT_m}-P \label{the first max profit of edge}
\end{eqnarray}

Thus, by substituting (\ref{unique 1 Nash}) into (\ref{the first max profit of edge}), and let $a=(R+TR)e^{-\gamma zT_m}$, the additional profit maximization of the edge server is simplified as:
\begin{eqnarray}
\mathop{maximize}\limits_{P>0}  \Delta U_e=a(1-\sqrt{\frac{X\varphi_1}{Pe^{-\gamma zT_m}}}) \label{the first profit of edge}
\end{eqnarray}

{\bf Theorem 3.} Under the same computer power of miners, the Edge Server can achieve the optimal profit.
\par  $proof.$ From (\ref{the first profit of edge}), We proof that the first derivative of $\Delta U_e$ with respect to the expected fee of miners P is positive and the second derivative of $\Delta U_e$ with respect to P is negative as following.

\begin{eqnarray}
\frac{\partial\Delta U_e }{\partial P}=\frac{1}{2}\sqrt{X\varphi_1 e^{-\gamma zT_m}}P^{-\frac{3}{2}}>0 \label{the first derivative of edge I}
\end{eqnarray}
and 
\begin{eqnarray}
\frac{\partial^2\Delta U_e }{\partial P^2}=-\frac{3}{4}\sqrt{X\varphi_1 e^{-\gamma zT_m}}P^{-\frac{5}{2}}<0 \label{the second derivative of edge I}
\end{eqnarray}

From (\ref{the first derivative of edge I}) and (\ref{the second derivative of edge I}), $\Delta U_e$ is strictly concavity that can be satisfied. Therefore, the EMG $E_m$ can obtain the optimal profit with the situation for the same computer power of miners. The process of proof has done.

\subsection{the different computer power for each miner}

\emph{1) Stage \uppercase\expandafter{\romannumeral2}: Miner's Game}: in this situation, the edge server and mobile devices are the miners for mining. The additional benefits for the edge server is represented as:

\begin{eqnarray}
\Delta U_e=(R+TR)\frac{x_i}{\sum\limits_{j \in \mathcal M}x_j}e^{-\gamma zT_m}-\sum\limits_{i \in \mathcal M}p_ix_i \label{the second max profit of edge}
\end{eqnarray}
and the utility function of miner $i$ is represented as:
\begin{eqnarray}
U_m=p_i\frac{x_i}{\sum\limits_{j \in \mathcal M}x_j}e^{-\gamma zT_m}-\varphi_2 x_i.\label{the second utility of miner i}
\end{eqnarray}

{\bf Theorem 4.} There is a Nash equilibrium in $E_m=\{E,M,U_m\}$

\par $proof.$ For computer power space of each miner $x_i$, it is a continuous and non-empty value range. In order to proof the concavity of $U_m$ in (\ref{the second utility of miner i}), we calculate the first derivative of $U_m$ and the second derivative of $U_m$ with respect to $x_i$. The process is as following:

\begin{eqnarray}
\frac{\partial U_m}{\partial x_i}=P_ie^{-\gamma zT_m}\frac{\partial\alpha}{\partial x_i}-\varphi_2 
\end{eqnarray}
and 
\begin{eqnarray}
\frac{\partial^2 U_m}{\partial x_i^2}=P_ie^{-\gamma zT_m}\frac{\partial^2\alpha}{\partial x_i^2}
\end{eqnarray}
where
\begin{eqnarray}
\frac{\partial \alpha}{\partial x_i}= \frac{\sum\limits_{i \ne j}x_i}{(\sum\limits_{j \in \mathcal M}x_j)^2}>0
\end{eqnarray}
and
\begin{eqnarray}
\frac{\partial^2 \alpha}{\partial x_i^2}= -2\frac{\sum\limits_{i \ne j}x_i}{(\sum\limits_{j \in \mathcal M}x_j)^3}<0
\end{eqnarray}

Therefore, $U_m$ satisfies the concave condition with respect to $x_i$. According to \cite{han2012game}, the Nash equilibrium is in the EMG $E_m$. Let (\ref{the second utility of miner i})=0, we can get the best response function of miners as following.

\begin{eqnarray}
\frac{\partial U_m}{\partial x_i}=p_ie^{-\gamma zT_m} \frac{\partial\alpha}{\partial x_i}-\varphi_2 =0 \label{the first derivative of utility function miner i}
\end{eqnarray}

\begin{eqnarray}
x_i^*=F(x)=\sqrt{\frac{p_ie^{-\gamma zT_m}\sum\limits_{i \ne j}x_j}{\varphi_2}}-\sum\limits_{i \ne j}x_j \label{the second optimal x_i}  
\end{eqnarray}

{\bf Theorem 5.} There is a unique Nash equilibrium in EMG $E_m$ if the following condition is satisfied

\begin{eqnarray}
\frac{2(M-1)}{p_i}<\sum\limits_{j \in \mathcal M}\frac{1}{p_j} \label{condition 2}
\end{eqnarray}

\par Let $x_i^*$ represent the Nash equilibrium of the EMG. If $x_i^*=F(x)$ satisfies the three condition of the standard function shown in Definition 1, then the EMG exists a unique Nash equilibrium \cite{han2012game}. 

Firstly, for the positivity, we need to prove $\sum\limits_{i \ne j}x_j<\frac{p_ie^{-\gamma zT_m}}{\varphi_2}$. Let (\ref{the sum miner i}) minus (\ref{the second Nash for miner i}), we have
\begin{eqnarray}
\sum\limits_{i \ne j}x_j=\frac{\varphi_2}{p_ie^{-\gamma zT_m}}(\frac{M-1}{\sum\limits_{j \in \mathcal M}\frac{\varphi_2}{p_ie^{-\gamma zT_m}}})^2
\end{eqnarray}

Under the condition of (\ref{condition 2}), we have
\begin{eqnarray}
\sum\limits_{i \ne j}x_j<\frac{p_ie^{-\gamma zT_m}}{4\varphi_2} \label{the sum x_i > 0}
\end{eqnarray}

It is easily to obtain  $\sum\limits_{i \ne j}x_j<\frac{p_ie^{-\gamma zT_m}}{\varphi_2}$. Thus, it satisfies the positivity.

Secondly, for the Monotonicity, we calculate $F(x')-F(x)$ under the condition $x' \ge x$. It is similar with (13). We have (32).

\begin{figure*}[t]
	\begin{align*}
    F(x')-F(x)= (\sqrt{\frac{Pe^{-\gamma zT_m}}{\varphi_2}}-\sqrt{\sum\limits_{i \ne j}x_j'}-\sqrt{\sum\limits_{i \ne j}x_j})(\sqrt{\sum\limits_{i \ne j}x_j'}-\sqrt{\sum\limits_{i \ne j}x_j})
    \tag{32}
    \end{align*}
    \begin{align*}
    \lambda F(x_i)-F(\lambda x_i)= (\lambda -\sqrt{\lambda})\sqrt{\frac{Pe^{-\gamma zT_m}\sum\limits_{i \ne j}x_j}{\varphi_2}}
    \tag{33}
    \end{align*} 
    \hrulefill 
\end{figure*}
Obiously, $\sqrt{\sum\limits_{i \ne j}x_j'}-\sqrt{\sum\limits_{i \ne j}x_j}>0$, and we have

\begin{eqnarray}
\setcounter{equation}{34}
\sqrt{\frac{p_ie^{-\gamma zT_m}}{\varphi_2}}-\sqrt{\sum\limits_{i \ne j}x_j'}-\sqrt{\sum\limits_{i \ne j}x_j} \in \notag\\ (\sqrt{\frac{p_ie^{-\gamma zT_m}}{\varphi_2}}-2\sqrt{\sum\limits_{i \ne j}x_j'}, \notag\\ \sqrt{\frac{p_ie^{-\gamma zT_m}}{\varphi_2}}-2\sqrt{\sum\limits_{i \ne j}x_j})
\end{eqnarray}

From (\ref{the sum x_i > 0}), we can obtian $\sqrt{\frac{p_ie^{-\gamma zT_m}}{\varphi_2}}-2\sqrt{\sum\limits_{i \ne j}x_j}>0$. Thus, it is obviously exist the condition that satisfies $F(x')-F(x)>0$. The proof is now completed.

Finally, for condition(3) of Definition 1. We prove the formula of $\lambda F(x_i)>F(\lambda x_i)$ in the condition as $\lambda>1$. From (33), we can obtain it.

{\bf Theorem 6.} In EMG $E_m$, the unique Nash equilibrium solution of computing power for a miner $i$ is shown by
 
\begin{eqnarray}
x_i^*=\frac{M-1}{\sum\limits_{j \in \mathcal M}\frac{\varphi_2}{p_ie^{-\gamma zT_m}}}-\frac{\varphi_2}{p_ie^{-\gamma zT_m}}(\frac{M-1}{\sum\limits_{j \in \mathcal M}\frac{\varphi_2}{p_ie^{-\gamma zT_m}}})^2 \label{the second Nash for miner i}   
\end{eqnarray}

$proof.$ From (\ref{the first derivative of utility function miner i}), we obtain

\begin{eqnarray}
\frac{\sum\limits_{i \ne j}x_j}{(\sum\limits_{j \in \mathcal M}x_j)^2}=\frac{\varphi_2}{p_ie^{-\gamma zT_m}}
\end{eqnarray}

Then, for all the miners, we calculate the sum of the above formula as following:

\begin{eqnarray}
(M-1)\frac{\sum\limits_{i \ne j}x_j}{(\sum\limits_{j \in \mathcal M}x_j)^2}=\sum\limits_{j \in \mathcal M}\frac{\varphi_2}{p_ie^{-\gamma zT_m}}
\end{eqnarray}

Thus, we can get
\begin{eqnarray}
\sum\limits_{j \in \mathcal M}x_j=\frac{M-1}{\sum\limits_{j \in \mathcal M}\frac{\varphi_2}{p_ie^{-\gamma zT_m}}} \label{the sum miner i}
\end{eqnarray}

From (\ref{the second optimal x_i}), we have 
\begin{eqnarray}
\sum\limits_{j \in \mathcal M}x_j=\sqrt{\frac{p_ie^{-\gamma zT_m}\sum\limits_{i \ne j}x_j}{\varphi_2}}
\end{eqnarray}

which means
\begin{eqnarray}
\sum\limits_{j \in \mathcal M}x_j=\sqrt{\frac{p_ie^{-\gamma zT_m}(\sum\limits_{j \in M}x_j-x_i)}{\varphi_2}} \label{another expressed the sum miner i}
\end{eqnarray}

By substituting (\ref{the sum miner i}) to (\ref{another expressed the sum miner i}), we have

\begin{eqnarray}
\frac{M-1}{\sum\limits_{j \in \mathcal M}\frac{\varphi_2}{p_ie^{-\gamma zT_m}}}=\sqrt{\frac{p_ie^{-\gamma zT_m}}{\varphi_2}(\frac{M-1}{\sum\limits_{j \in \mathcal M}\frac{\varphi_2}{p_ie^{-\gamma zT_m}}}-x_i)}
\end{eqnarray}

We will get the Nash equilibrium solution of computing power for miner $i$ with mathematical operation. The process of proof has done.

\par Therefore, we can adopt the solution for obtaining the Nash equilibrium for the edge server. We will analyze the optimal profit in Stage \uppercase\expandafter{\romannumeral1}.

\emph{2) Stage \uppercase\expandafter{\romannumeral1}: Edge Server's optimal profit}: It is similar to \ref{The same computer power for each miner}. What we need to analyze is the additional benefits of the computing power from the edge server recruiting mobile miner $i$. The computer power provided by miner $i$ is $x_i$. Therefore, the additional benefits for the edge server represented as:

\begin{eqnarray}
\Delta U_e=(R+TR)\frac{x_i}{\sum\limits_{j \in \mathcal M}x_j}e^{-\gamma zT_m}-p_i \label{the second additional profit of edge}
\end{eqnarray}

Thus, by substituting (\ref{the second Nash for miner i}) and (\ref{the sum miner i}) into (\ref{the second additional profit of edge}), and let $a=(R+TR)e^{-\gamma zT_m}$, the additional profit maximization of the edge server is simplified as:

\begin{eqnarray}
\mathop{maximize}\limits_{p_i>0} \Delta U_e=a(1-\frac{M-1}{p_i\sum\limits_{j \in \mathcal M}\frac{1}{p_j}}) \label{the second profit of edge}
\end{eqnarray}

{\bf Theorem 7.} Under the different computer power of each miner $i$, the Edge Server can achieve the optimal profit.

$proof.$ From (\ref{the second profit of edge}), we can calculate the first derivative of $\Delta U_e$ with respect to the expected fee $p_i$ of miner $i$ is positive and that of the second derivative of $\Delta U_e$ is negative as following:

\begin{eqnarray}
\frac{\partial \Delta U_e}{\partial p_i}=\frac{a(M-1)\sum\limits_{i \ne j}\frac{1}{p_j}}{(p_i\sum\limits_{j \in \mathcal M}\frac{1}{p_j})^2}>0 \label{the first derivative of edge II}
\end{eqnarray}

and

\begin{eqnarray}
\frac{\partial^2 \Delta U_e}{\partial p_i^2}=\frac{-2a(M-1)(\sum\limits_{i \ne j}\frac{1}{p_j})^2}{(p_i\sum\limits_{j \in \mathcal M}\frac{1}{p_j})^3}<0 \label{the second derivative of edge II}
\end{eqnarray}
\par \quad
From (\ref{the first derivative of edge II}) and (\ref{the second derivative of edge II}), $\Delta U_e$ is strictly concavity that can be satisfied. Therefore, the EMG $E_m$ can obtain the optimal profit with the situation for the different computer power of miners. The process of proof has done.

\par \quad
\par \quad
\section{Simulation Results}\label{simlation-section}
 
We will evaluate the performance of the incentivizing mechanism in EMG proposed in this paper. We consider an edge server that recruits a set of mobile devices to mine in mobile blockchain system. We adopt algorithm I to achieve the unique Stackelberg Equilibrium solution for miners. Then the edge server maximizes its profit by substituting the expected fee of miners. For a given expected fee imposed by EMG, the following sub game is solved first. The optimal response of follower game is substituted into leader game, and the optimal price is obtained. Similar algorithms can also be used for the situation of different fee provided by the edge server. We set 1000 blocks adopting Node.js\cite{js2016node}. To simplify the experiment, the size of block is the same and there are 10 transactions exist in each block to mine.

\begin{algorithm}[t]
	\caption{The Algorithm for finding Stackelberg Equilibrium under the same expected fee} 
	\hspace*{0.02in} {\bf Input:}
	initial expected fee $p_0$, parameter $\theta$ , precision threshold $\varepsilon$\\
	\hspace*{0.02in} {\bf Output:} 
	the optimal expected fee $p_{opt}$, the computing power of miner $c_i$, the count of successful mining $\mathcal S$
	\begin{algorithmic}[1]
		
		\State{\bf Initialization:} 
		\State Set initial $p_0$, parameter $\theta \in (0,1)$, precision threshold $\varepsilon > 0$\\
		Let $X_1, \cdots, X_n$ be $n$ independent random 0-1 variables, where $X_i$ takes 1 with the successful mining of miner $i$.
		\State {\bf Repeat:}
		\State Each miner $i$ provides its computing power $c_i$ to mine
		\State If the mining is successful, $X_i$=1
		\State $p_{i+1}=p_i(1+\theta)$
		\State $i \gets i+1$
		\State {\bf Until:}
		\State $p_{i+1}-p_i < 0$ and $\frac{|p_{i+1}-p_i|}{p_i} < \varepsilon$
		\State $\mathcal S = \sum\limits_{i=1}^n X_i$
		\State {\bf Return:}
		\State $p_{opt}=p_i$, $c_i$, $\mathcal S$
	\end{algorithmic}
\end{algorithm}

Firstly, we evaluate the probability of successful mining versus the computing power of edge server. The computing power from the edge server is the critical factor that decides the probability of successfully mining block, as shown in Fig 1.

\begin{figure}[htp]
	\centering
	\includegraphics[width=3.0in,height=3.0in,clip,keepaspectratio]{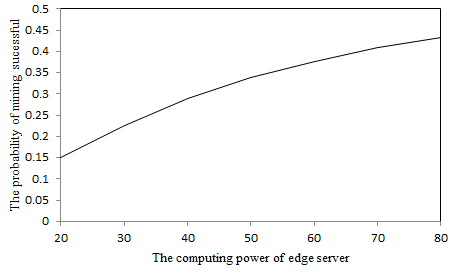}
	\caption{\small The probability of mining successful vs The computing power of the edge server}
\end{figure}

We secondly investigate the optimal expected fee for mobile devices under the same computing power versus the fixed reward R by blockchain system.

\begin{figure}[htp]
	\centering
	\includegraphics[width=3.0in,height=3.0in,clip,keepaspectratio]{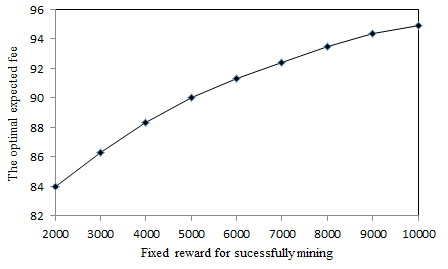}
	\caption{\small The optimal expected fee vs Fixed reward for successfully mining}
\end{figure}

\par The Fig 2 shows that the optimal expected fee for mobile devices increase with the increasing of the fixed reward R. The fixed reward is higher, the expected fee of the edge server is higher. Miners are more willing to respond to the recruitment of edge servers to provide higher computing power to mine. Therefore, the edge server can provide a higher expected fee to recruit mobile devices and obtain higher computing power to improve the probability of mining success, and thus obtain greater profit. Thus, with the increase of fixed rewards provided by blockchain system, the individual computing power provided by each miner also increases.

\par We next evaluate two situations for the profit of Edge Server versus the different computing power of recruited miners and the different computing power provided by edge server under the same total computing power for them. The first one is that the computing power from edge server is set to 50 constantly, and when the computing power of this recruited mobile devices increases, the reward curve of the edge server changes. Another is that the total computing power of mobile miners is set to 50, and when the computing power from edge server increases, the reward curve of the edge server changes accordingly. From Fig 3, we find that the profit of the edge server increase rapidly with the increase of recruited mobile devices in the initial stage. That’s because, in the initial stage, with the mobile devices recruited by the edge server, the computing power is increased rapidly, and the probability of mining success is increased, so the reward is increased. However, with the addition of more mobile devices, the increasing probability of mining success is not obvious, but the edge server still has to pay the corresponding remuneration, so the cost increases sharply, and the total income increases slowly. In addition, under the different computing power for each mobile device, the profit of the edge server increases with the increase of total computing power compared with that under the same computing power of mobile devices. However, Fig 4 shows that when the computing power from edge server accounts for a larger proportion of the total computing power, there is little difference between the benefits of the edge server under the two situations. That’s because the edge server is dominant in that case, and the computing power from mobile device is little and makes little contribution to mining.

\begin{figure}[htp]
	\centering
	\includegraphics[width=3.0in,height=3.0in,clip,keepaspectratio]{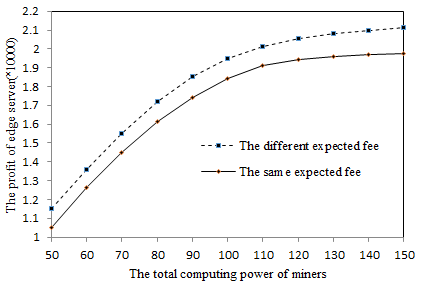}
	\caption{\small The profit of the edge server vs The total computing power of miners}
\end{figure}

\begin{figure}[htp]
	\centering
	\includegraphics[width=3.0in,height=3.0in,clip,keepaspectratio]{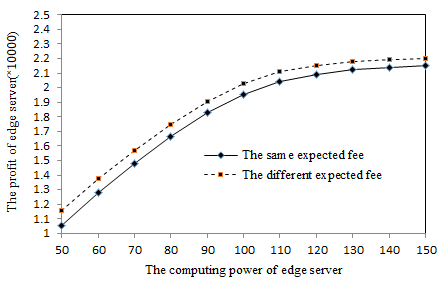}
	\caption{\small The profit of the edge server vs The computing power of the edge server}
\end{figure}

\par Finally, we compare the profit of edge server in EMG with MDG\cite{xiong2018optimal}. For the EMG, we consider that the computing power of the edge server accounts for 10\%, 50\% and 90\% of the total computing power respectively. In addition, the total computing power that provided by all miners in two schemes is the same. We investigate the profit of the edge server versus the total computing power. From Fig 5 and Fig 6, we can find that the profit of the edge server in both schemes increases with the increase of the total computing power. However, the profit of the scheme proposed in this because the edge server itself participates in mining, reducing the information transmission delay and improving the efficiency. In the EMG method, the edge server does not participate in mining, and all mining tasks are assigned to the mobile device, which will generate a large communication cost and a large delay, so it reduces the probability of mining success and reduces the profit of the edge server

\begin{figure}[htp]
	\centering
	\includegraphics[width=3.0in,height=3.0in,clip,keepaspectratio]{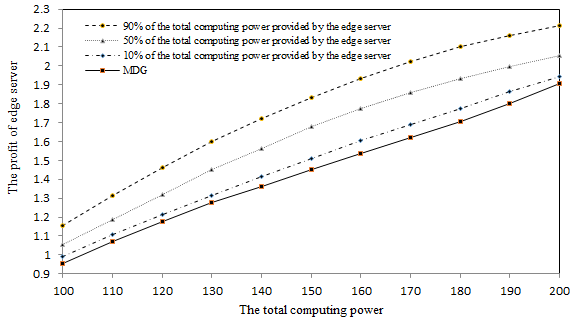}
	\caption{\small The profit of the edge server vs The total computing power under the different computing power provided by miners}
\end{figure}

\begin{figure}[htp]
	\centering
	\includegraphics[width=3.0in,height=3.0in,clip,keepaspectratio]{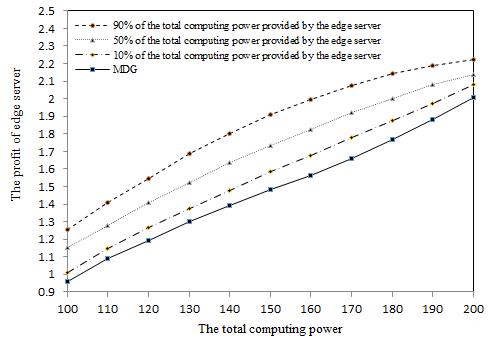}
	\caption{\small The profit of the edge server vs The total computing power under the same computing power provided by miners}
\end{figure}

\section{Conclusion}

In this paper, we discuss the challenges of mobile blockchain in mining. In order to address the challenges of high mining cost for mobile devices, this paper proposes an incentive mechanism based on edge computing. We developed a two-stage Stackelberg Game model to jointly optimize the reward of edge servers and recruited mobile devices. The edge server recruits mobile devices to compete with it in mining. The edge server sets the expected fee according to the computing power of the mobile device, and the mobile device provides the computing power for the edge server. We prove that there is a unique Nash Equilibrium solution in this game under the same expected fee or different expected fee. In addition, we show the result curve of the profit for the edge server with the different ratio between the computing power from edge server and mobile devices. The results show that the contribution of the computing power from the edge server to its profit is more than that of it recruited mobile devices under the same condition. In addition, the proposed scheme has been compared with the MDG scheme for the profit of the edge server. The results show that the profit of the proposed scheme is more than that of the MDG scheme under the same conditions.

\section*{Acknowledgment}

This work is supported by the National Science Foundation of China(No.61662039), Science and technology project of Jiangxi Provincial Department of Education (No. GJJ170967), Jiangxi Key Natural Science Foundation (No. 20192ACBL20031), Project of Teaching Reform in Jiujiang University(No. XJJGYB-19-47).

\ifCLASSOPTIONcaptionsoff
  \newpage
\fi



%

 \bibliographystyle{unsrt}
\bibliography{mybibliography}

\end{document}